\documentclass[%
superscriptaddress,
longbibliography,
preprint,
amsmath,amssymb,
aps,
prx,
]{revtex4-1}

\usepackage{array}
\usepackage{color}
\usepackage{graphicx}
\usepackage{dcolumn}
\usepackage{bm}
\usepackage{amsmath, amssymb}
\usepackage{makecell}
\usepackage{multirow}
\usepackage{adjustbox}
\usepackage{float}
\usepackage{verbatim}

\usepackage[mathlines]{lineno}

\graphicspath{{figures/}}

\newcommand{\bs}[1]{\boldsymbol{#1}}

\begin{document}

\title{Thermodynamically Informed Multimodal Learning of High-Dimensional Free Energy Models in Molecular Coarse Graining}

\author{Blake R. Duschatko}
\email{bduschatko@g.harvard.edu}
\affiliation{John A. Paulson School of Engineering and Applied
Sciences, Harvard University, Cambridge, MA 02138, USA}

\author{Xiang Fu}
\affiliation{Massachusetts Institute of Technology, Cambridge, MA 02138, USA}

\author{Cameron Owen}
\affiliation{Department of Chemistry and Chemical Biology, Harvard University, Cambridge, MA 02138, USA}

\author{Yu Xie}
\affiliation{John A. Paulson School of Engineering and Applied
Sciences, Harvard University, Cambridge, MA 02138, USA}

\author{Albert Musaelian}
\affiliation{John A. Paulson School of Engineering and Applied
Sciences, Harvard University, Cambridge, MA 02138, USA}

\author{Tommi Jaakkola}
\affiliation{Massachusetts Institute of Technology, Cambridge, MA 02138, USA}

\author{Boris Kozinsky}
\email{bkoz@seas.harvard.edu}
\affiliation{John A. Paulson School of Engineering and Applied
Sciences, Harvard University, Cambridge, MA 02138, USA}
\affiliation{Robert Bosch LLC Research and Technology Center, Watertown, MA 02472, USA}

\begin{abstract}    

We present a differentiable formalism for learning free energies that is capable of capturing arbitrarily complex model dependencies on coarse-grained coordinates and finite-temperature response to variation of general system parameters. 
This is done by endowing models with explicit dependence on temperature and parameters and by exploiting exact differential thermodynamic relationships between the free energy, ensemble averages, and response properties. Formally, we derive an approach for learning high-dimensional cumulant generating functions using statistical estimates of their derivatives, which are observable cumulants of the underlying random variable. The proposed formalism opens ways to resolve several outstanding challenges in bottom-up molecular coarse graining dealing with multiple minima and state dependence. This is realized by using additional differential relationships in the loss function to significantly improve the learning of free energies, while exactly preserving the Boltzmann distribution governing the corresponding fine-grain all-atom system. 
As an example, we go beyond the standard force-matching procedure to demonstrate how leveraging the thermodynamic relationship between free energy and values of ensemble averaged all-atom potential energy improves the learning efficiency and accuracy of the free energy model. The result is significantly better sampling statistics of structural distribution functions. The theoretical framework presented here is demonstrated via implementations in both kernel-based and neural network machine learning regression methods and opens new ways to train accurate machine learning models for studying thermodynamic and response properties of complex molecular systems.

\end{abstract}

\maketitle

\clearpage 



\section{Introduction}

Molecular dynamics (MD) is a vital tool in computational materials design that allows one to probe the thermodynamic and kinetic properties of matter at atomistic resolution. Of central importance to MD is the accuracy of the force fields used to capture interactions between atoms. While classical empirical force fields derived from experimental observation have dominated the field for many decades, due to the prohibitive computational burden of \textit{ab initio} methods, progress in state-of-the-art machine learning methods has helped to increase the speed and size of simulations while reaching \textit{ab initio} accuracy~\cite{ace, gap, schnet, nequip, allegro, flare}. These developments have opened unprecedented capabilities in the modeling of phase transitions, surface reconstruction, catalytic reactions, and biomolecular conformational changes~\cite{yu-sic, anders-micron, david-catalysis, flarepp, cameron-catalysis, cameron-gold, niti, muller2021mlff-review, yan2023mlff-applications,noe202proteins,unke2020reactive,linfeng2020hundredmillion}. 

Existing atomistic machine learning force field (MLFF) methods have been shown to very effective at regression of the atomistic potential energy surface (PES) of the reference quantum mechanical model. However, because they include all of the atomistic degrees of freedom (DOFs), they require small integration timesteps to resolve fast atomic motion, while accurately computing forces on all DOFs. This limitation is particularly significant for describing soft matter phenomena characterized by a wide range of time scales, such as in dynamics of liquid crystal ordering, polymer reptation, and protein conformation. Coarse graining (CG) methods have long been used to address this issue by integrating out fast DOFs, allowing for larger timesteps and fewer force calculations. Most widely used are coarse grained force fields (CGFFs) with simple fixed functional forms parameterized "top-down" to reproduce experimental observations~\cite{oplsua,martini,cabs,unres,protein_models}, while "bottom-up" CGFFs are derived from a fixed microscopic atomistic model~\cite{mscg,rel_entropy}. The latter approach is appealing since there is a rigorous way to maintain thermodynamic consistency of the statistical ensembles of the two scales of description. In the bottom-up CG model, the effective energy function of the coarse-grained coordinates is the potential of mean force (PMF)~\cite{mscg,rel_entropy,ibi}, or the free energy of the constrained CG system, which accounts for the entropy of the all-atom (AA) fine-grain configurations. 

Several approaches have been proposed for bottom-up coarse graining. For example, the iterative Boltzmann inversion (IBI) method is a relatively simple approach that allows models to self-consistently target radial distribution functions (RDF) of particular systems \cite{ibi, multistate-ibi}. Alternatively, relative entropy minimization (REM)~\cite{rel_entropy, deep_rel_entropy} has proven to aid in the task of learning accurate structural distributions that correctly capture correlated behavior. However, both approaches suffer from various limitations. In particular, the IBI approach is limited in its ability to reproduce accurate structural correlations outside of the RDF it is trained on, while the REM approach requires comparatively expensive training due to the need of multiple molecular dynamics (MD) simulations to self-consistently converge the potential.
It is also possible, however, to learn the PMF by regressing its gradients to match Boltzmann averages of forces on the CG sites~\cite{mscg,flarecg}. Compared to IBI methods, these force-matching approaches are capable of capturing a wider array of structural and thermodynamic properties while being significantly faster to train than REM methods. 

Despite their appeal, bottom-up CG approaches often require complex functional forms to capture many body interactions. As such, machine learning (ML) approaches have emerged as a promising resolution to this issue~\cite{clementi_mlcg, gap_cg, clementi_gnn, autoencoders, flarecg,deep_rel_entropy}. These ML CGFFs have been shown to be capable of accurate modeling of protein hydration free energies~\cite{clementi_mlcg}, joint learning of interactions and inverse CG to AA mappings~\cite{autoencoders}, as well as on-the-fly learning and transferability across structural spaces~\cite{flarecg}. Moreover, ML approaches present a more promising means of accurately capturing higher-order structural correlations that had previously been limited by the simple functional forms of interactions~\cite{deep_rel_entropy}. 

In the context of the current state-of-the-art in ML CGFF methods, and motivated by the recognized limitations in the field~\cite{jin_cgPerspective}, we propose a method that opens new research avenues for exploring some of the outstanding problems in coarse graining. Our approach is based on Sobolev training and leverages previously unused information of the potential energy means within a unified framework, which exhibits significantly improved learning efficiency and accuracy. 

The first challenge is thermodynamic representability, or correct description by the CG model of thermodynamic properties such as pressures, potential energies, or entropies, in accord with the AA reference. Existing methods rely on learning thermodynamic properties independently from the free energy functions, such as learning separately the mean potential energy ~\cite{dual_potential}, and therefore do not enforce exact consistency between the AA and CG models.
The method formulated herein offers a rigorous, consistent way of learning the PMF and other thermodynamic properties subject to exact constraints.

Another capability missing in existing CGFF models is multimodal learning, or the ability of the model to efficiently utilize multiple types of available training data from AA simulations. Current bottom-up coarse graining methods typically only use average AA forces for training, and this results in poor sample efficiency and accuracy of the free energy ~\cite{flarecg,deep_rel_entropy,iterative-ygb}. While some previous models were designed to learn from additional thermodynamic properties such as densities~\cite{mscg-npt,pressure-match}, the PMF remains fixed in the learning procedure as additional variables are incorporated and is hence not improved. In contrast, our proposed formalism can efficiently utilize multiple types of available training data, readily available from fine-grained AA sampling simulations, in a manner that simultaneously improves the accuracy of the PMF and related thermodynamic properties, due to the consistency enforced among them. 

Lastly, we note that our new methodology introduces a new direction in exploring a wide range of response properties and the possibility of consistently simulating systems under the influence of external fields. Notably, both AA and CG models have lacked this capability so far. Our approach addresses this challenge by endowing differentiable models with arbitrary parameter inputs that can be used to learn and predict CG-level response properties of any order. Our work extends recent works on Sobolev learning of generalized  ground-state response properties ~\cite{stefano-polarization, fieldschnet} to the case of free energies with arbitrary parameter and temperature dependence.

We present the framework in general terms and subsequently focus on a representative example to illustrate its benefit. Specifically, we include the mean AA potential energies as an additional previously unused learning target for CG free energies. We show that such energy-informed CG models more accurately capture free energy differences and interaction correlations with far less data, specifically in scenarios with multiple free energy minima. We demonstrate these principles first on a simple model system using a kernel-based approach and further with an equivariant neural network CG representation of free energy of hexane, highlighting how accurately complex structural correlations are preserved, and confirming the utility of our formalism. 

\section{Results}

\subsection{Unified Framework for Thermodynamically Informed Differential Models}

Let us consider a system with microscopic atomistic (AA) potential energy $U(\bs{r}, \lambda_a)$ that depends in a general nonlinear way on the set of $n$ atomic coordinates $\bs{r} \equiv \{r_i \} \in \mathbb{R}^{3n}$, and arbitrary parameters $\lambda_a$. The parameters $\lambda_a$ can be local or global and include the Bravais lattice vectors (or volume) and any generalized forces, such as electric field, magnetic field, electrostatic or chemical potential, etc. The conjugate properties $\partial U/\partial \lambda_a$ are correspondingly stress (or pressure), polarization,  magnetization, charge or particle number. This generalization can be applied to lower-dimensional coarse grained representations that can be used to accelerate calculations of thermodynamic properties. This coarse-graining is achieved by integrating over $\bs{r}$, the $n$ coordinate degrees of freedom (DOFs) of the microscopic AA system, at a particular value of inverse temperature $\beta = 1/k_BT$ to obtain the free energy $W(\bs{R}, \lambda_a, \beta)$ of the coarse-grained (CG) system as a function of parameters $\lambda_a$ and the $N$ CG DOF coordinates $\bs{R} \equiv \{R_I \}\in \mathbb{R}^{3N}$ defined by the $\mathbb{R}^{3n} \to \mathbb{R}^{3N}$ mapping ${R}_I = {M}_I(\bs{r})$ for each CG DOF $I$. In macroscopic thermodynamics, all atomistic DOFs are integrated out fully, with the free energy depending only on global parameters such as volume and temperature. In the context of molecular coarse-graining, the set of coordinates $\bs{R}$ typically correspond to positions of CG beads, where the map  and the free energy $W$ is referred to as the potential of mean force (PMF).
We consider the latter case in this work, but the formalism generally applies to any type of dimensionality reduction, including the context of collective variables for enhanced sampling.
We implement our differentiable thermodynamics framework in the context of bottom-up coarse graining, capitalizing on its rigorous statistical consistency between the AA and the CG models.
Thermodynamic consistency requires that the partition function, and therefore statistical weights, are preserved, thereby defining the free energy $W$.
\begin{equation}\label{eq:partitionf}
    Z = e^{-\beta W(\bs{R}, \lambda_a, \beta)} = \int d^n\bs{r} e^{-\beta U(\bs{r}, \lambda_a)}\delta^N({M_I}(\bs{r}) - {R_I}) 
\end{equation}
The function $W(\bs{R}, \lambda_a, \beta)$ determines all thermodynamic and response properties of the CG system, and it is our goal to learn it, assuming we have the knowledge of the underlying AA potential energy function $U(\bs{r}, \lambda_a)$. Herein we follow the common choice of a linear mapping of AA to CG coordinates ${M_I}(\bs{r})$ that defines the coordinate of a CG unit in terms of a collection of AA coordinates ${R}_I = {M}_I(\bs{r}) = \sum_j c_{Ij}{r}_j$, where $c_{Ij}$ are the mapping coefficients between AA coordinates and the CG unit $I$. 

Directly using Eq. \ref{eq:partitionf} to compute the value of $W$ for every CG configuration is well known to be intractable due to the high dimensionality of the integral. In a class of CG methods referred to as force matching, the potential of mean force (PMF) is learned using its derivatives with respect to CG coordinates ${R}_I$ with a loss function consisting of the mean squared residual between the all-atom instantaneous forces on CG sites and the CG forces that are gradients of the PMF~\cite{mscg}. Our approach generalizes this approach by including in the learning objective an expanded set of derivatives the CG free energy function $W$. To this end, we expand the dimensionless free energy $\beta W = -\ln Z$ with a Taylor series of its parameters and use its various differential coefficients in the learning task.
\begin{align}\label{eq:taylor}
    \beta W(\bs{R} + \Delta \bs{R}, \lambda_a + \Delta \lambda_a, \beta+ \Delta \beta) &= \beta W(\bs{R}, \lambda_a, \beta) + \frac{\partial (\beta W)}{\partial\beta} \Delta \beta + \beta\frac{\partial W}{\partial R_I}\Delta R_I + \beta\frac{\partial  W}{\partial \lambda_a}\Delta \lambda_a \nonumber \\
    &+ \frac {1}{2} \beta\frac{\partial^2  W}{\partial \lambda_a\partial \lambda_b}\Delta \lambda_a\Delta \lambda_b + \frac {1}{2} \frac{\partial^2 (\beta W)}{\partial \lambda_a\partial \beta}\Delta \lambda_a \Delta\beta + \cdots
\end{align}
where we use the summation convention. The key to our approach is that derivatives of the free energy, which are the coefficients of this expansion, are ensemble averages of the system's response properties that are readily obtained from microscopic constrained dynamics or Monte-Carlo sampling computations driven by the known AA energy function $U(\bs{r}, \lambda_a)$.

Similarly to training conventional atomistic MLFFs settings, one approach to fitting the free energy is to train via force labels. The mean force is an ensemble average of atomistic forces constrained by each CG configuration, given by
\begin{equation}\label{eq:mean-force}
    {F}_I(\bs{R}) = -\frac{\partial W}{\partial R_I} = - \left< \frac{\partial U}{\partial R_I} \right> _{\bs{R}, \beta} = \frac{1}{Z} \int d\bs{r}^n \Big(\sum_{i\in I} {f}_i\Big)e^{-\beta U(\bs{r})}\delta^N({R}_I - {M}_I(\bs{r}))
\end{equation}
where $f_i$ are AA forces obtained from AA simulations using constrained sampling, and the ensemble average is taken only over AA configurations $r_i$ that map to CG coordinates $R_I$. If $\tilde W$ is the learned model of the true free energy $W$, the commonly used force-matching loss function takes the form
\begin{equation}\label{eq:mean-force-loss}
    \mathcal L_{MF} = \sum_t^{T}\sum_I^N \Big| {F}_I(\bs{R}_{t}) + \frac{\partial \tilde W(\bs{R}_t)}{\partial{R}_{I,t}} \Big|^2 
\end{equation}
where $t$ indexes the timeframe of a given training configuration, $T$ is the number of CG resolution training frames, $I$ indexes the CG coordinate, $N$ is the number of CG degrees if freedom. We discuss in Methods alternative formulations of the loss function that do not require constrained AA dynamics for training label generation.

The fully general expression of Eq. \ref{eq:taylor} indicates many possibilities to utilize additional properties for training. For instance, a derivative with respect to a parameter $\lambda_a$ is the linear response coefficient
\begin{equation}\label{eq:pmf_response}
    \frac{\partial \beta W}{\partial \lambda_a} = \beta \left< \frac{\partial U(\bs{r}, \lambda_a))}{\partial \lambda_a} \right>_{\bs{R}, \beta}
\end{equation}
that relates to the pressure of the CG system if the $\lambda_a$ is the volume, or polarization if $\lambda_a$ is the electric field.
 Other training targets are made clear from the fact that derivatives of the dimensionless free energy $\beta W$ with respect to $\beta$ are the statistical cumulants of the microscopic potential energy $U$, since $\beta W$ is the cumulant generating function for the potential energy as the underlying random variable
\begin{equation}
\frac{\partial }{\partial \beta}\beta W(\bs{R}, \lambda_a, \beta) = \left< U(\bs{r}, \lambda_a) \right>
\label{eq:pmf_U_mean}
\end{equation}

\begin{equation}
\frac{\partial^2 }{\partial \beta^2}\beta W( \bs{R}, \lambda_a,\beta) = \left< (U-\left< U\right>)^2 \right> = Var(U)
\label{eq:pmf_U_var}
\end{equation}
We can also include mixed derivatives related to temperature dependent response quantities of the CG system, noting that the ensemble average operation does not commute with the differentiation operation with respect to parameters of the PMF, e.g.
\begin{equation}
\frac{\partial^2 \beta W}{\partial \lambda_a \partial \lambda_b} = \beta \left< \frac{\partial^2 U}{\partial \lambda_a \partial \lambda_b}\right> - \beta^2  \ Cov \left(\frac{\partial U}{\partial \lambda_a} ,\frac{\partial U}{\partial \lambda_b} \right)
\label{eq:pmf_covar}
\end{equation}

\begin{equation}
\frac{\partial^2 \beta W}{\partial \lambda_a \partial \beta} = \left< \frac{\partial U}{\partial \lambda_a }\right> - \beta  \ Cov \left( U ,\frac{\partial U}{\partial \lambda_a} \right)
\label{eq:pmf_covar_mixed}
\end{equation}

\begin{figure}[h!]
    \centering
    \includegraphics[scale=0.9]{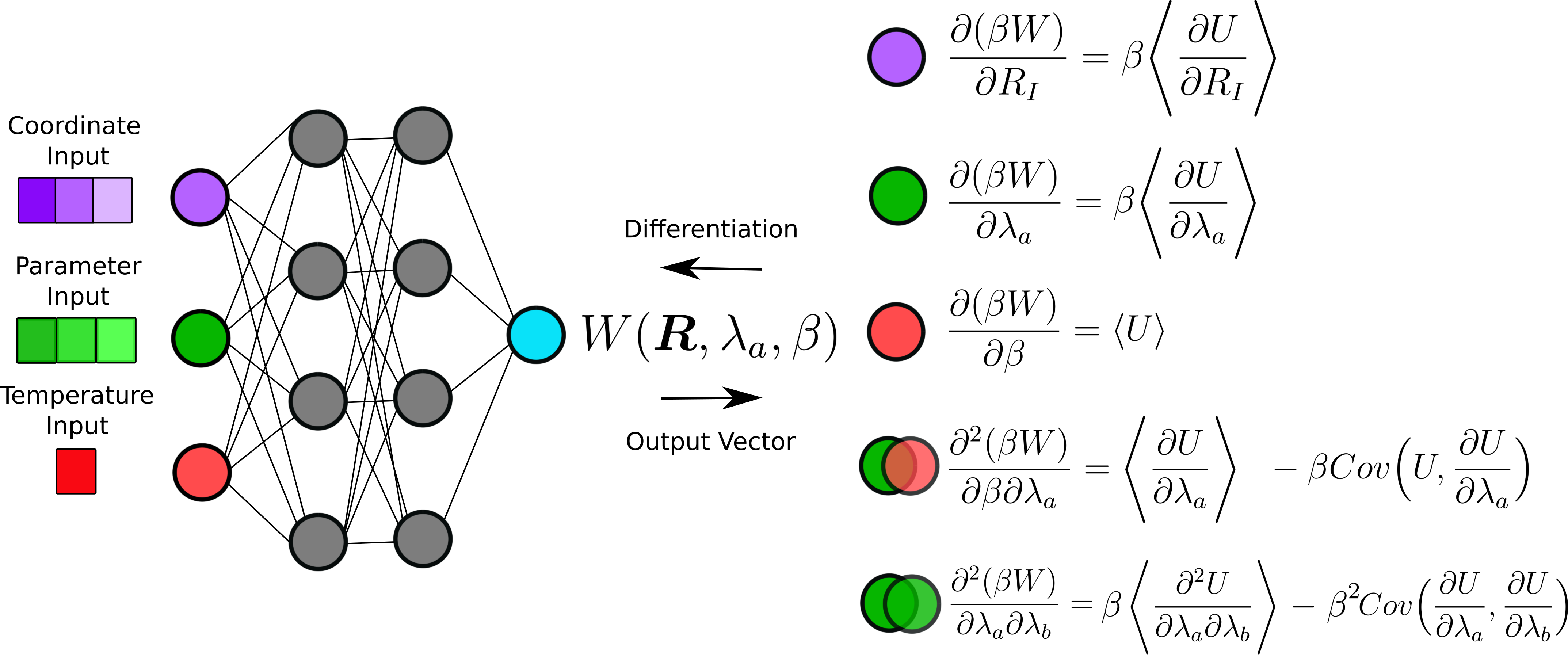}
    \caption{\textbf{Thermodynamically Informed Neural Network Framework for Free Energy Models.} In addition to system coordinates (purple), new inputs can be introduced into machine learning models, such as temperature (red) or other global parameters (green), including external fields. The resulting free energy output can be differentiated with respect to these parameters, giving access to new observables and field responses at any order.}
    \label{fig:framework}
\end{figure}

The scheme, which we call a "thermodynamically informed neural network", is depicted in Fig.~\ref{fig:framework} and applies generally to other methods such as Gaussian process regression. This differentiability enables access to an extensive range of previously inaccessible physical quantities, both for prediction and for model training. We derive inspiration from the idea of Sobolev training~\cite{sobolev-pinns, sobolev-elastoplasticity,Czarnecki2017SobolevTF}, which is beneficial if the training data for the derivatives is computationally cheap to obtain, which is the case for molecular simulations. Learning can be accomplished by including various combinations of available labels and target properties into the loss function 
\begin{align}\label{eq:meanforce+u}
    \mathcal L_{constr} = \sum_{m}\gamma_m\sum_t^{N_t} \left| \mathcal D_m \left[\beta W(\bs{R}_t,\lambda_a, \beta)\right] - \left< {\mathcal {K}_m} \Big[\beta U(\bs{r}, \lambda_a)\Big] \right>_{\bs{R}_t, \beta}\right|^2
\end{align}
where  $\mathcal D_m$ are the various derivative operators acting on $\beta W$ on the left hand sides of equations above, $\mathcal {K}_m$ are the operators in the ensemble averages on the right hand sides, and $\gamma_m$ are weights for the various loss terms. The sums in the loss function run over the number of training frames $N_T$ and the number of CG sites $N$.
Crucially, when physical quantities are obtained as derivatives of the free energy, they exactly satisfy the correct physical symmetries and conservation laws. This is analogous to ensuring energy conservation when forces are learned as gradients of the energy in standard AA MLFFs. Such conservation laws are not enforced exactly if physical quantities are learned directly with separate dedicated models.
An immediate implication for molecular coarse graining is that these differential relations provide many more possibilities to train a thermodynamically consistent CG free energy model (Eq. \ref{eq:taylor}) in a bottom-up fashion. The training labels for the absolute value of the free energy is intractable to obtain due to the difficulty in summing over all states; however, derivatives corresponding to response properties (e.g. Eqs. \ref{eq:pmf_response}, \ref{eq:pmf_covar}), are readily obtained from microscopic sampling simulations and can help learn parameter dependent properties, such as stress and polarization, for the finite-temperature CG system. Even more directly important is the ability provided by Eqs. \ref{eq:pmf_U_mean} and \ref{eq:pmf_U_var} to use AA potential energy cumulants for learning the PMF. This addresses the significant difficulty of predicting relative free energies, particularly in free energy landscapes with multiple minima. Because force-matching traditionally relies on force training data alone (Eq. \ref{eq:mean-force}), it can be exceedingly difficult to capture relative free energies of the minima as well as of the large-barrier transition states. In the limit that we can use for training the force labels over a dense set of Cartesian coordinates of the CG sites, at fixed $\beta$, the PMF is determined up to a constant (since only gradients are used). In practice, however, there are often gaps in the training set due to rare occurrence of some configurations. Therefore,  the ability to train PMF models using AA potential energies is very valuable. Implementation of this formalism simply requires that the PMF models be made explicitly dependent on temperature so that appropriate derivatives can be taken and used in the loss function. In the following, we provide two demonstrations of the value of our formalism and using additional training labels: one utilizing the loss function in Eq.~\ref{eq:meanforce-noisy+u} using the Allegro model~\cite{allegro}, and another using the sparse Gaussian process (SGP) CG framework based on FLARE~\cite{flarecg}.

\subsection{Gaussian Process Regression for a Model System}
We first illustrate the proposed augmented free energy training by combining force-matching Eq. \ref{eq:mean-force} with mean AA potential energy Eqs. \ref{eq:pmf_U_mean} within a Gaussian process (GP) regression context.
The concept of a loss function in GPs is well understood and discussed in detail in literature~\cite{bishop}. To employ these ideas in the sparse Gaussian process (SGP) approach, we must additionally define covariance relationships to jointly train against and predict both forces and mean potential energies. Specifically, we use the framework introduced in our earlier work based on the FLARE framework ~\cite{flarecg} wherein the PMF is written as a sum of local contributions as 
\begin{equation}
    W(\bs{R}, \lambda_a, \beta) = \sum_I^N w_I(\bs{R},\lambda_a, \beta)
\end{equation}
The covariance between local free energy contributions, $w$, is given in this case by a dot product kernel function between two descriptors
\begin{equation}
    \text{cov}(w_I(\bs{d}_I), w_J(\bs{d}_J)) = k(\bs{d}_I, \bs{d}_J) = \sigma^2\Big(\bs{d}_I\cdot\bs{d}_J\Big)^\xi
\end{equation}
where $\xi$ is the kernel power, $\sigma$ the signal hyper parameter, and the descriptors $\bs d_I(\bs R, \lambda_a, \beta)$ are functions of the coordinates, parameters, and temperature. The detail of their implementation are given in Methods. 
The bi-linearity of the covariance allows us to specify the kernel elements between local free energies and total mean potential energies as
\begin{equation}
    \text{cov}(\langle U \rangle, w_J(\bs{d}_J)) = \frac{\partial}{\partial\beta}\Big(\beta\sum_I k(\bs{d}_I, \bs{d}_J)\Big)
\end{equation}
where $I$ labels each CG site. Expressions for the covariance between mean potential energies and forces can be similarly derived. By incorporating the new properties into the kernel of the SGP, one gains not only the ability to train against additional targets, but also a means of predicting principled quantitative uncertainties on predictions of these properties, which have proven to be quite useful in previous work~\cite{flarecg, flarepp}.


We first illustrate the utility of the SGP approach by considering a simple low-dimensional model previously used to examine coarse graining strategies ~\cite{clementi_mlcg}. The fine-grain potential energy is given by
\begin{equation}\label{eq:toy}
    \beta V(x,y) = \frac 1{50}(x-4)(x-2)(x+2)(x+3) + \frac{y^2}{20} + \frac{\sin(3(x+5)(y-6))}{25} + \Big(\frac{x}{2}\Big)^3\Big(\frac{y}{\sqrt{170}}\Big)^2
\end{equation}
The CG coordinate is taken to be the value $x$, with integration over $y$ serving as the coarse graining. Integration for computing the free energy, as well as the mean fine grained potential energy, is performed numerically, and the coefficients are chosen to provide numerically stable solutions. For this example we set $\beta=1$.

In the following, we demonstrate that learning against average fine-grain potential energies in addition to forces improves the PMF model accuracy, and verify that the use of more complex and expressive temperature-dependent descriptors are not the sole reason for improved performance.
To examine the impact of including the mean potential energy in the kernel on helping to capture relative free energy differences, we consider models in two regimes, one in a large data regime and the other in a low data regime. The model free energy introduced in Eq.~\ref{eq:toy} has two energy minima separated by an energy barrier. The SGP's are trained with data collected only from these two minima. In the low data regime, two training points are randomly sampled from each basin, while in the large data regime we sample four (with a total of 8). We further define three types of models: a) models that have no temperature dependence in the descriptor and hence no energy information in the kernel matrix, b) models that have temperature dependence but no energy information in the kernel matrix, trained only on forces, and c) have both temperature dependent descriptors and energy-dependent kernels, trained on both force and energy labels. For the sake of brevity, we will from here on denote models with energy-dependent kernels as ``energy labeled models."

\begin{figure}
    \centering
    \includegraphics{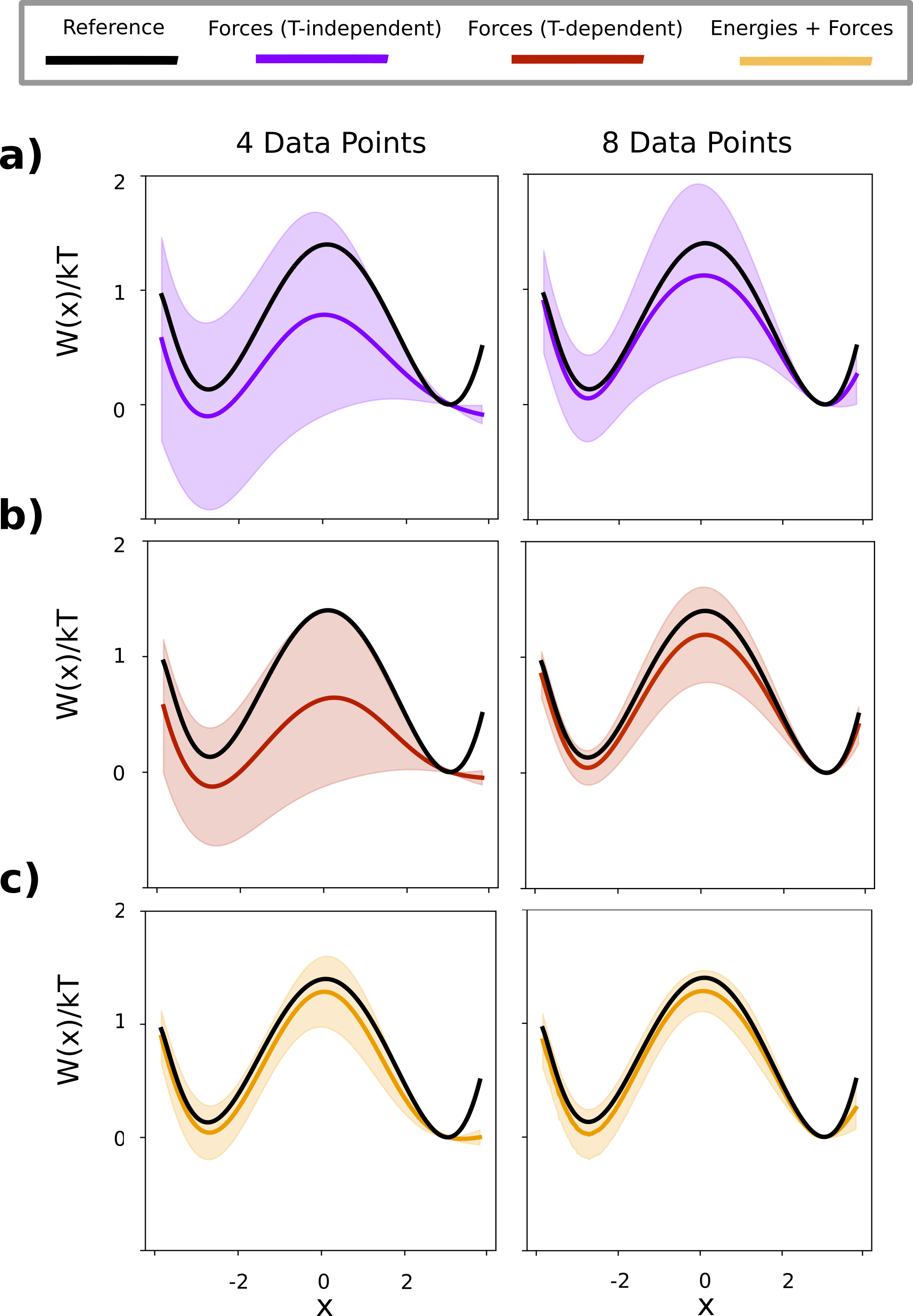}
    \caption{\textbf{A simple model of potential-energy informed learning.} Shown are the mean (solid lines) and standard deviations (shaded regions) of ensembles of 10 models in two different regimes. In column 1, models contain 4 data points, with 2 sampled from each basin. In column 2, models contain 8 data points, with 4 sampled from each basin. a) The mean and standard deviation of models with force only data and temperature-independent descriptors. b) The mean and standard deviation of models with force only data and temperature-dependent descriptors. c) The mean and standard deviation of models with force and energy data, as well as temperature-dependent descriptors.}
    \label{fig:toy}
\end{figure}

The results, depicted in Fig.~\ref{fig:toy}, make clear the benefit of using fine-grained energy labels for training the PMF. We note that we are only interested in the relative energy differences between the basins, and in this figure the zero of the free energy is taken to be at the right basin minimum. As such, the models are made to agree at this location. In the low-data regime, depicted in the first panel of subplots a-c, the energy labeled models capture relative energy difference with much higher accuracy. Moreover, we observe that energy labeled models are additionally more certain in domains outside of the training set, as depicted by the standard deviation of the model predictions. Further, we note the improvement of the temperature dependent model in the absence of energy labels in relation to its temperature independent counterpart. The increased expressiveness of the temperature dependent descriptors allows for a more descriptive model, thereby improving the achievable accuracy. Nonetheless, it is clear that the energy labels are the most significant source of improvement on the attainable accuracy compared to the force-only models. 

\subsection{Equivariant Neural CGFFs for Small Molecules}

\begin{figure*}[h!]
    \centering
    \includegraphics[scale=.9]{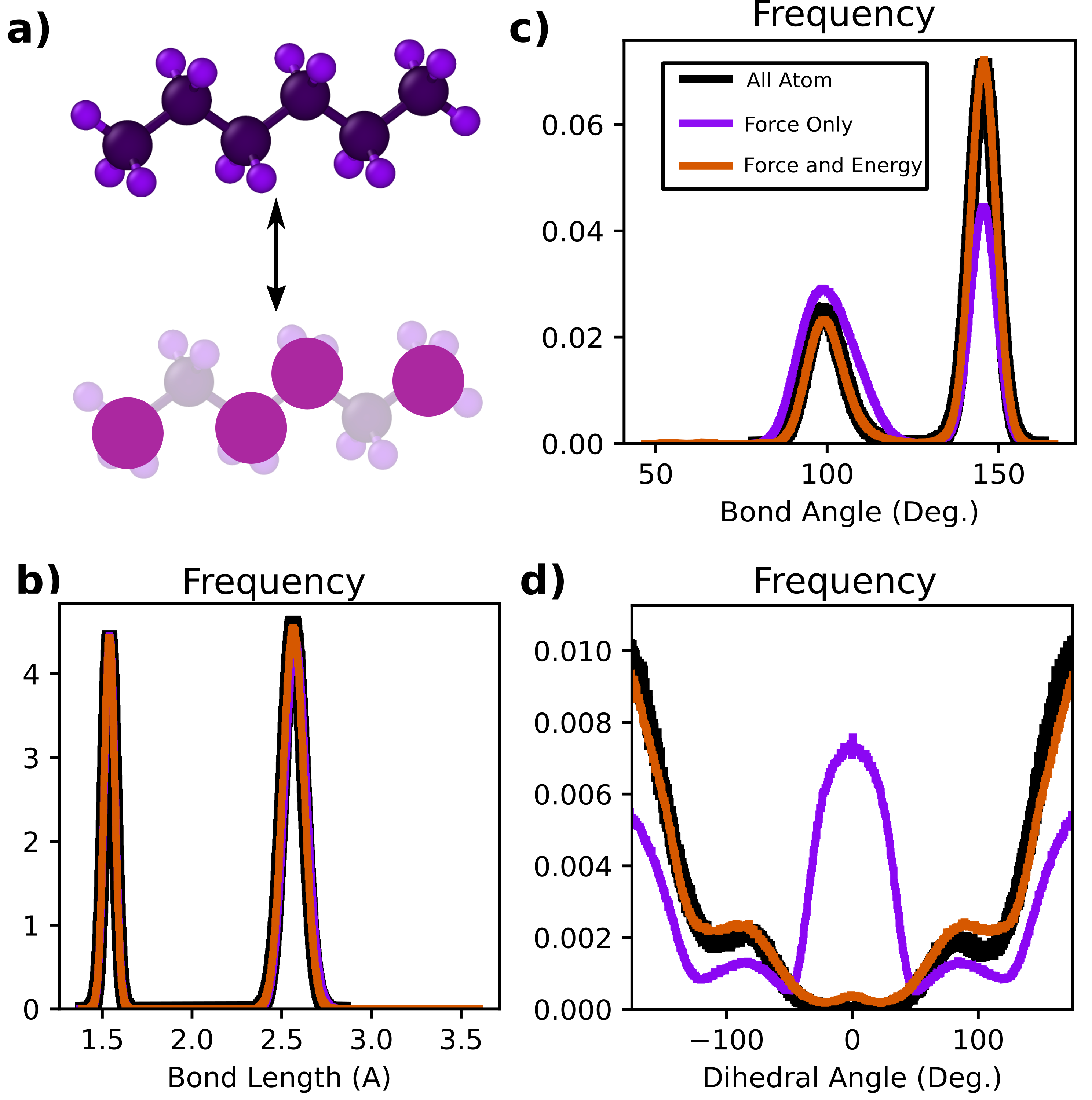}
    \caption{\textbf{Comparison of Structural Distributions With and Without Potential Energies}. a) A hexane molecule is mapped to its most interior and exterior carbons for a 4 site CG model. b) the bond length distribution of energy-labeled models (orange) and force-only models (purple) compared to the all-atom baseline (black). c) the bond angle distribution between the two pairs of three consecutive CG sites. d) the dihedral angle distribution for the four CG sites.}
    \label{fig:hexane-dists}
\end{figure*}

As a more realistic test case, we shift our attention to the coarse graining of small molecules, relevant in the practical context of liquid solvents, or the design of low-dimensional PMFs for implicitly solvated small proteins. To illustrate the approach in the context of neural network CGFFs, we apply our approach to a coarse grained representation of a single hexane molecule. Hydrocarbons have been frequently explored in the CG methods and applications literature ~\cite{iterative-ygb, flarecg, mscg-liquids, iterative-mscg, mb-correlations}. In particular, capturing relative energy differences and accurate sampling of correlated interactions in these molecules has been identified as an important outstanding challenge~\cite{flarecg, iterative-ygb, iterative-mscg}. To explore the impact of our temperature-dependent energy-informed approach on the learning of structural correlations in real molecules, we consider a 4-site AA to CG mapping of hexane, depicted in Fig.~\ref{fig:hexane-dists}a. 
We use this case to demonstrate our approach, and specifically the loss function proposed in Eq.~\ref{eq:meanforce-noisy+u}, with an equivariant neural network model of the free energy, based on the Allegro architecture~\cite{allegro}.
We examine the performance of two model types, those with and without training on AA potential energy labels. As metrics of model accuracy, we compare bond length, bond angle, and dihedral angle distributions, obtained with CG and reference AA data. Specifically, the bond-length distribution is defined as the pairwise distance between nearest-neighbor CG sites shown in Fig.~\ref{fig:hexane-dists}a, the bond angle distribution is between the first three and last three CG sites in the molecule, and the dihedral distribution is for the single dihedral angle in the coarse grained molecule. Our models are trained on 200,000 time frames (configurations) of AA data obtained from NVT simulations at 250K using the OPLS force field~\cite{opls}.

As seen in Fig.~\ref{fig:hexane-dists}b-d, the models containing energy labels are far superior. In particular, Fig.~\ref{fig:hexane-dists}c and d show that the bond angle and dihedral angle distributions of the energy labeled model match the all-atom baseline with much greater fidelity. Further, we see in Fig.~\ref{fig:hexane-dists}d that the high-energy states near zero degrees, which in the AA model is very sparsely sampled, are significantly oversampled by the non-energy labeled CG model.

We note that the two Allegro models, both with and without energy labels, are trained in a relatively low data regime compared to other CG NN models in the literature developed for small molecules. For example, CG NN models of alanine dipeptide have in previous works required upwards of a million frames of AA data to train~\cite{clementi_mlcg}. Equivariant models like Allegro and NequIP have been widely shown to have higher data efficiency in learning compared to other NNs~\cite{allegro,nequip,voth-water}.  At the same time, in the data limited regime, the addition of energetic information leads to even higher data efficiency and much more accurate models. 

\begin{figure*}[h!]    
    \centering
    \includegraphics[scale=.75]{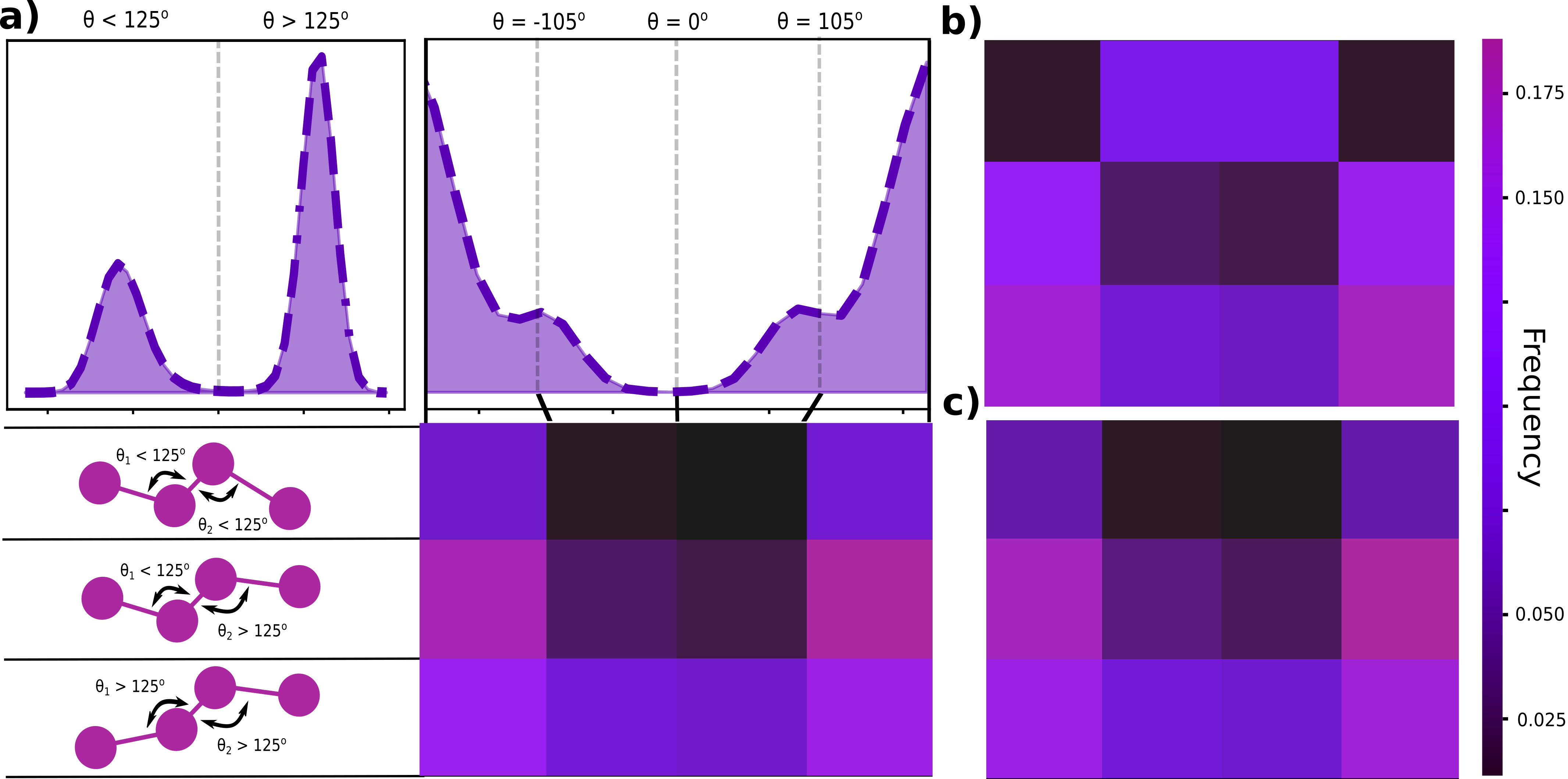}
    \caption{\textbf{Comparison of Structural Correlations With and Without Potential Energies}. a) All-atom angular distribution of the molecule corresponding to the CG mapping is shown, with two peaks on either side of 125$^{\circ}$. The CG molecule can adopt three different angular conformations corresponding to whether each of its two angles is above or below 125$^{\circ}$. The dihedral angle distribution of the all-atom model is also shown and separated into four distinct regions. The relative sampling of the 12 unique correlated dihedral states as measured by an all-atom MD simulation is shown. b) Relative sampling of an optimized Allegro model that is temperature dependent but has seen no energy labels. f) Relative sampling of an optimized Allegro model that is temperature dependent and has seen both force and energy labels. a-c share the color bar shown.}
    \label{fig:hexane-fes}
\end{figure*}

It is well known that force-matching based CG models can miss key features in the relative values of correlated structural distributions~\cite{generalized-ygb, iterative-ygb, andrienko-correlations, iterative-mscg, mb-correlations}. To this end, we additionally examine the free energy surface (FES) of the hexane molecule, defined as follows. We define the FES in this case to describe the relative probabilities of the 12 distinguishable unique structural states of the molecule at CG resolution. By choosing a 4-site CG mapping in Fig.~\ref{fig:hexane-dists}a, we partition the bond-angle and dihedral angle distributions into 3 and 4 domains, respectively. This is illustrated in Fig.~\ref{fig:hexane-fes}a. The sampling of these states for a baseline OPLS AA model is depicted in Fig.~\ref{fig:hexane-fes}a. Further supporting the results in Fig.~\ref{fig:hexane-dists}, we see that the FES generated by the energy-labeled CG PMF model (Fig.~\ref{fig:hexane-fes}c) is significantly more accurate than the force-only model (Fig.~\ref{fig:hexane-fes}b).

The totality of these improvements is summarized and further quantified in Table~\ref{table:hexane} by considering a variety of error metrics. Moreover, we demonstrate the improved learning rate of energy-labeled models by comparing the performance of models with 100,000 and 200,000 data frames. We note that the noisy validation force loss in Eq.~\ref{eq:meanforce-noisy+u} of the Methods is effectively the same across models. This further emphasizes the fact that mean force errors are not a sufficient metric of quality of models for systems with distinct minima states, in problems where occupancy distributions are of interest. In all other error metrics shown with respect to the FES and intra-molecular distributions, the energy labeled models perform better. Specifically, the mean absolute error (MAE) values in all structural distributions are lower for the energy labeled models. Further, we consider which of the 12 FES bins, depicted in Fig.~\ref{fig:hexane-fes}, have the highest error and lowest error. In both cases, these two error metrics are higher when energy labels are absent. In addition, the energy-labeled models having been trained on only 100,000 frames of data are more accurate than even the force-only models in the 200,000 data frame regime. In summary, the results on the hexane system indicate that already just one additional training target of our proposed approach, the energy labeling using mean AA potential energies, improves many facets of the PMF learning process with already available AA data, with no additional computational burden, and even requires fewer training data overall.

\section{Discussion}

The results of this work establish a rigorous theoretical framework for improved learning of high-dimensional free energy models, such as the PMF in the context of coarse graining. 

A primary goal of coarse graining approaches lies in the accurate estimation of free energies as functions of coordinates of reduced degrees of freedom. Estimation of free energy functions is a long-standing challenge in statistical mechanics. In abstract statistical sense, our work provides a general framework for constructing differentiable models to learn high-dimensional cumulant generating functions. In the conceptual vein of Sobolev training, we train the models on derivatives of the target function, which are the cumulants of the underlying random variables, that we obtain by statistical sampling. 
We note the distinction from the idea of physically informed neural networks (PINNs)~\cite{pinns}, where differential expressions involving only the output are used as additional regularization terms in the loss function. Instead, we augment the model inputs with parameter inputs and use exact differential relations to allow for additional training targets, improving the model transferability, accuracy and training efficiency.

Specifically in the context of molecular coarse graining, we provide a way to learn the free energy by using physically observable thermodynamic and response properties in a unified thermodynamically consistent manner by identifying these statistical cumulants with 
 derivatives of the free energy. The advantage of this approach is the simplicity of its implementation, where existing CG models are endowed with addition explicit inputs on parameters, specifically including the temperature and modifying the loss function with matching observable - derivative pairs. This is the first method to date, to our knowledge, that utilizes potential energies for training in such a way and that enables the values (not only gradients) of the model PMF to be directly improved through the new training information. 

Further, the resulting models require no additional computational overhead, since atomistic constrained dynamics simulations produce trajectories that contain force training labels along with total potential energies, from which a variety of statistical cumulants can be estimated at no additional computational cost. This framework can be seamlessly implemented into existing learning architectures, and is shown to produce more accurate PMFs with a higher learning efficiency. As demonstrations of the improvements at fixed temperature, we have considered a low-dimensional toy model example, as well as a realistic molecular coarse graining procedure for the hexane molecule. 

We note that additional computational overhead may be incurred in neural network implementations of the Sobolev training strategy, as a result of needing multiple passes of backpropagation for higher order derivatives. However, we expect that the statistical sampling of AA labels will dominate the computational cost nonetheless. An issue to be resolved for unconstrained sampling is that the loss function over noisy labels in Eq.~\ref{eq:meanforce-noisy+u} of Methods is easy to implement only up to first derivatives (our demonstration includes the use of forces and potential energies). If using the variance of the potential energy (Eq. \ref{eq:pmf_U_var}), it is necessary to estimate the mean potential energy for each CG configuration, prior to formulating a noisy unconstrained estimation of the variance. Our formalism applies straightforwardly at any order when using constrained sampling to estimate ensemble averages, but efficient implementation of such estimates using noisy unconstrained labels for higher order derivatives should be addressed in future work. 

As an extension of this multi-modal learning, we recognize that our method also offers a promising avenue for improving the state point transferability of CG models. For example, while CG models have been trained before against multi-temperature force data with an explicit temperature input parameter~\cite{rafa_temp}, our proposed inclusion of cumulants of the AA potential energy data in the training labels brings significant additional information via the PMF's temperature dependence via its gradients. As a result, this is a promising new means of simulating systems across different phases with improved model transferability. More generally, our model formulation allows for the learning of the PMF dependence on any thermodynamic state point directly through an expanded set of ensemble observables. Demonstration of this capability is left to a future investigation.

We note some limitations of the present analysis as opportunities for future work. For example, we should expect that, in the case of potential energy labeling at fixed temperature and volume, the energy labels will be maximally helpful to the free energy when the all-atom constrained entropic contribution is close to zero, i.e. when the atomistic potential energy labels are most informative. While any additional information from the energy labels will help, the extent to which energies help in learning the PMF should be more closely explored, particularly for large systems where the total potential energy label may contain limited information.

Broadly, for a given mapping between the fine and coarse grained representations, we present a unified differential framework for connecting the two scales in a rigorous consistent way and for capturing arbitrarily complex nonlinear and coupled dependencies of coarse-grained free energy on system parameters. Our method can be readily generalized to include additional thermodynamic observables as targets, such as stresses, polarization, and magnetization. This can be accomplished by including as inputs to the CG model and differentiating it with respect to variables such as strain, electric field, and magnetic field, respectively, and matching in the loss function with the corresponding AA observables. This framework enables the development of coarse grained temperature-dependent models in the presence of any set of generalized forces acting in a variety of thermodynamic ensembles. Examples of such future work might include exploring temperature and pressure-dependent phase transitions, coupled electro-mechanial response, and extensions to other thermodynamic ensembles such as the grand canonical ensemble. Examples to be explored include higher-order derivatives, corresponding to learning from and predicting e.g. heat capacity and compressibility. In general, this framework will enable a better understanding of what aspects of coarse-grained models can be improved by training with different combinations of all-atom ensemble averages.


\section{Methods}

The loss function discussed in the main text targeting constrained dynamics labels is not the only choice of a thermodynamically consistent loss function. In place of fitting to these constrained force labels as in Eq. \ref{eq:mean-force-loss}, it is common in force-matching schemes to minimize a noisy loss function using unconstrained estimates of the forces on CG units. In this case, the framework proposed in this work for including arbitrary thermodynamic properties can be extended to these schemes using the relationships between free energy and other thermodynamic properties, via the the loss function
\begin{align}\label{eq:meanforce-noisy+u}
    \mathcal L_{noisy} = \sum_{m}\gamma_m\sum_t^{N_t} \left| \mathcal D_m \left[\beta W(\bs{M}(\bs{r}_t),\lambda_a, \beta)\right] -  {\mathcal {K}_m} \Big[\beta U(\bs{r}_t, \lambda_a)\Big] \right|^2
\end{align}
where the frame index $t$ now indexes instantaneous atomistic frames, and not constrained CG values. We emphasize the subtle but important difference in this expression compared to Eq.~\ref{eq:mean-force-loss}: The loss function in Eq.~\ref{eq:mean-force-loss} regresses the properties of the PMF to their thermodynamic averages, while Eq.~\ref{eq:meanforce-noisy+u}  regresses the properties of the PMF to the noisy, instantaneous atomistic predictions of the property. It is not immediately obvious that the two loss functions produce the same minimum. We provide a derivation of this equivalence in the SI. 

The implementation of this loss function requires the model to be explicitly dependent on temperature. To accomplish this, we modify the Allegro architecture to introduce the temperature as part of the input node features. The original Allegro model uses one-hot embedding of the atom type as the node feature: $\mathbf{h}_i = \mathrm{1Hot}(Z_i)$, where $Z_i$ is the discrete type of node $i$ (chemical species in all-atom potentials). We augment $\mathbf{h}_i$ by concatenating a temperature embedding with a Gaussian basis: $\mathbf{h}_i = \mathrm{1Hot}(Z_i)\ \|\ B(\beta)$, where $\|$ denotes concatenation, $B$ is the Gaussian basis function, and $\beta = 1/k_B T$.  A similar encoding was described in an earlier work\cite{rafa_temp}, which however did not consider the use of thermodynamic differential relationships to augment the training targets, as proposed here. In our case, we compute the derivative
${\partial \tilde W(\mathbf{R}^N, \beta)}/{\partial \beta}$ through backpropagation and use the result in a generalized loss function.

Similarly, Gaussian process models also require temperature dependence in order to leverage additional training targets. For the Gaussian process model, we choose to define descriptors as modified variants of those introduced in our earlier work~\cite{flarecg} based on the Atomic Cluster Expansion (ACE)~\cite{ace}. Specifically, in order to leverage thermodynamic relationships of the PMF, we introduce temperature dependence into the SGP by adding temperature dependent embedding functions in the descriptors. The non-temperature dependent SGPs are constructed in the FLARE MLFF with a descriptor of the form
\begin{equation}\label{eq:ace}
    d_{is_1s_2n_1n_2\ell} = \sum_m c_{is_1n_1\ell m}c_{is_2n_2\ell m}
\end{equation}
where the ACE atomic base is given by
\begin{equation*}
    c_{isn\ell m}(\bs{R}) = \sum_{j\in\rho_i} R_n(|\bs{R}_{ij}|) Y_\ell^m(\hat{\bs{R}}_{ij})\delta_{s,s_j}
\end{equation*}
while for the proposed temperature-dependent models, we introduce an expansion in a basis of functions of temperature
\begin{align*}
    c_{isn\ell m}(\bs{R}, \beta) = \sum_{\gamma}a^{n\ell}_\gamma\sum_{j\in\rho_i} R_n(|\bs{R}_{ij}|) Y_\ell^m(\hat{\bs{R}}_{ij})\delta_{s,s_j}\Gamma_{\gamma}(\beta)
\end{align*}
with new coefficients, $a$, that can couple to the radial and angular basis functions. In this work, we take the basis functions, $\Gamma$, to be the Chebyshev polynomials. With this change, the descriptor in Eq. \ref{eq:ace} becomes a temperature and parameter dependent descriptor such as those appearing in the main text. 

\subsection{Computational Details}

All production simulations were performed in the LAMMPS software package~\cite{lammps}, with atomistic models utilizing parameters from the OPLS force field~\cite{opls}. The parameter files were generated using the LigParGen server~\cite{ligpargen1,ligpargen2}. In order to train the Gaussian process models, a modified version of FLARE was developed to include new descriptors and kernels. For production CG simulations, custom LAMMPS software was developed for fast implementation of the NN and SGP models as pair styles. 

 In order to generate structural distributions for model validation, we run NVT simulations of each AA model for 2 nanoseconds, starting from 343 different initial velocities and equilibrating for 200 picoseconds. We use a timestep of $dt=1$ fs with a damping constant for a Nose-Hoover thermostate of $100*dt$. 

In the SGP models, we construct a set of 10 models having been trained on different random samplings of the data described in the main text. Each model is then optimized independently by minimizing the force mean squared error (MSE) between model predictions and the ground truth values over the full domain of the PMF shown in Fig.~\ref{fig:toy}. The ground truth values are computed via numerical integration of the PMF. Further details of the optimization procedure are provided in the SI.  

In the NN models, a temperature embedding was included in the architecture for both the force-only case and the case including atomistic potential energies. The energy dependent models were trained by optimizing a joint loss function as in Eq.~\ref{eq:meanforce-noisy+u} that includes a noisy force contribution as well as a noisy potential energy contribution. The relative weighting of these terms was taken to be a hyperparameter. The hyperparameters for each model type were chosen by minimizing the total validation loss of the models. In particular, the energy-labeled models included an energy-loss term in their validation metric, as in Eq.~\ref{eq:meanforce-noisy+u}. The details of parameter sweeps, as well as a sample input file, are expanded upon in the SI. 

\section{Data availability}
\noindent
All input and output files are available upon request. 

\section{Code availability}
\noindent 
All code is available upon request. 

\section{Acknowledgments}

The authors thank Anders Johansson, Stefano Falletta, and Steven Torrisi for helpful discussion. This work was supported by a NASA Space Technology Graduate Research Opportunity, under grant number 80NSSC20K1189, by the NSF through the Harvard University Materials Research Science and Engineering Center Grant No. DMR-2011754, and by a Multidisciplinary University Research Initiative sponsored by the Office of Naval Research, under Grant N00014-20-1-2418. All computational experiments utilized the resources provided and maintained by Harvard FAS Research Computing.

\section{Competing interests}
\noindent
The authors declare no competing financial or non-financial interests.

\section{Author Contributions}

B. R. D. implemented all of the new FLARE and LAMMPS code, carried out the proofs, and performed the computations.  X. F. provided an implementation of the method in the Allegro framework and preliminary computational experiments. B. K. conceived the approach and supervised the work. C. J. O. contributed to the conceptualization of figures and provided guidance on the training of FLARE models. Y. X. aided in the formulation and implementation of temperature-dependent ACE descriptors. A. M. provided support on the use of and development of Allegro models. T. J. reviewed the manuscript and supervised X. F.'s contribution to the project.  B. R. D. and B. K. wrote the manuscript, and all authors contributed to manuscript preparation. 

\clearpage

\bibliography{cg.bib}

\clearpage


\section{Figures and Tables}

\begin{table}[]
\renewcommand*{\arraystretch}{1.5}
\begin{tabular}{|c|cc|cc|}
\hline
Metric  & \multicolumn{2}{c|}{Force Only}  & \multicolumn{2}{c|}{Forces and Energies}    \\ \hline
Data   & \multicolumn{1}{c|}{100,000} & 200,000 & \multicolumn{1}{c|}{100,000} & 200,000 \\ \hline
Force Loss (meV/A)$^2$  & \multicolumn{1}{c|}{36.916}  & 36.869   & \multicolumn{1}{c|}{36.969}  & 36.867  \\ \hline
Bond Length MAE & \multicolumn{1}{c|}{0.0723}   & 0.009  & \multicolumn{1}{c|}{0.003}  & 0.004   \\ \hline
Bond Angle MAE  & \multicolumn{1}{c|}{0.0042}  & 0.0005   & \multicolumn{1}{c|}{0.0005}  & 0.0003  \\ \hline
Dihedral MAE  & \multicolumn{1}{c|}{0.0054}  & 0.003  & \multicolumn{1}{c|}{0.002}  & 0.0003  \\ \hline
FES MAE & \multicolumn{1}{c|}{0.0874}  & 0.035 & \multicolumn{1}{c|}{0.029}  & 0.006 \\ \hline
FES MAE Max & \multicolumn{1}{c|}{0.197}  & 0.088  & \multicolumn{1}{c|}{0.052}  & 0.016  \\ \hline
FES MAE Min & \multicolumn{1}{c|}{0.0121} & 0.001  & \multicolumn{1}{c|}{0.006}  & 0.001  \\ \hline
\end{tabular}
\caption{The table shows a variety of different error metrics for Allegro models, trained on hexane, both with and without energy labels. In both cases, results are given for low-data amounts (100,000 frames) and moderate data amounts (200,000 frames). Considered quantities include the noisy validation force loss given by Eq. \ref{eq:meanforce-noisy+u}, the MAE of the normalized bond length distribution, as well as the bond angle and dihedral angle distributions. In addition, we give the MAE of over the 12-basin free energy surface, including the maximum and minimum single-bin errors. }
\label{table:hexane}
\end{table}


\end{document}